\documentclass[prb,aps,12pt,a4paper,showpacs]{revtex4}
\usepackage{epsfig}
\begin{document}



\title{Generation of Correlated Binary Sequence from White Noise}

\author{F.~M.~Izrailev,$^1$ A.~A.~Krokhin,$^{1,2}$,
N.~M.~Makarov,$^{3}$}
\thanks{On sabbatical leave from Instituto de Ciencias,
        Universidad Aut\'{o}noma de Puebla,
        Priv. 17 Norte No. 3417, Col. San Miguel
        Hueyotlipan, Puebla, Pue., 72050, M\'{e}xico.}
\author{O.~V.~Usatenko,$^3$}

\affiliation{$^1$Instituto de F\'{\i}sica, Universidad Aut\'{o}noma de Puebla,\\
             Apartado Postal J-48, Puebla, Pue., 72570, M\'{e}xico}

\affiliation{$^2$Department of Physics, University of North Texas,
             P.O. Box 311427, Denton, TX 76203}

\affiliation{$^3$A.~Ya.~Usikov Institute for Radiophysics and
             Electronics,\\ Ukrainian Academy of Science,
             12 Proskura Street, 61085 Kharkov, Ukraine}

\date{\today}

\begin{abstract}

We suggest a method for generation of random binary sequences with
prescribed correlation properties. It is based on a kind of
modification of the widely used convolution method of constructing
continuous random processes. Apart from the theoretical interest,
this method can be used in various applications such as the design
of one-dimensional devices giving rise to selective transport
properties.

\end{abstract}


\maketitle

\section{Introduction}
Generators of random numbers or white-noise signals are customary
elements in modern digital electronics. Different algorithms are
used for this purpose. The quality of a generated white noise is
determined by the length of the sequence, elements of which can be
considered as uncorrelated. In many areas of physics such as the
engineering and signal processing, it is required to generate  a
colored noise, i.e. a correlated random process. Since the pair
correlations usually give the principal contribution to the
observable quantities, the problem of generation of a random
sequence with the prescribed {\it pair correlator} is of
particular interest. It has been known for a long time that
continuous colored noise with exponential correlations is
generated by the linear Ornstein –-Uhlenbeck process (see modern
review in Ref.~[\onlinecite{Hangi94}]). A more general method,
valid for the generation of continuous random sequences with an
{\it arbitrary} correlator, is based on the convolution of white
noise with the {\it modulation function} defined by the pair
correlator. Originally, the convolution method was proposed by
Rice.~\cite{Rice4454} Applications of some versions of this method
for the generation of random sequences with specific correlations,
including the long-range non-exponential correlations, can be
found in
Refs.~[\onlinecite{Saupe88Feder88,WOD95,Czir95,IzKr99,IzMak05,Grigo06}].

In the theory of spatially disordered systems the role of the pair
correlator (for random potentials) is emphasized by the fact that
it determines the scattering cross section in the Born
approximation (for weak potentials). As a result, many linear
transport characteristics (conductance, transmission and
reflection coefficients, localization length, etc.) are expressed
through the pair correlator.~\cite{Klyat,MakYur89FMYu90}

There are some examples of the systems (or processes) with
correlated disorder, for which the fluctuating parameter takes
discrete values. An example of such a system is the DNA
macromolecule. Here the genetic information is written using four
symbols that are the basic nucleotides. In the digital devices the
information is transmitted in a form of a telegraphic signal, i.e.
a binary code. The binary sequences is the limiting case of random
sequences with the least number of the basic elements. For
practical applications it is desirable to develop a method of
generation of a binary sequence with prescribed pair correlator --
a kind of colored noise containing two elements, e.g., "0" and
"1". Although there were some attempts to obtain a robust
algorithm for generation of a correlated binary sequence with the
purpose of increasing the performance of a pulse
radar,\cite{Polge} the problem is still lacking a general
solution. It is worth mentioning that there are methods of
generation of a correlated binary sequence, which are not based on
the properties of the correlation function, see, e.g., Ref.
[\onlinecite{Noncorr}]. It is not known yet, what are the
constraints (if there are any) for the pair correlator, imposed by
the fact that the sequence is {\it dichotomous} (binary). For
example, an attempt to generate a dichotomous sequence with the
correlations decaying according to the inverse power law, was
unsuccessful.\cite{CarpBerIvStanley-Nat02}

Recently, we addressed the mathematical problem of generation of a
dichotomous sequence with prescribed correlation
properties.~\cite{MUYaG05} We concentrated our attention on
statistical properties of binary \emph{additive} Markov chains. It
was shown that some special classes of correlations can be
reconstructed with the use of the so-called {\it memory function}.
The latter is related to the pair correlator through a quite
complicated linear integral equation.

In this paper we present a new approach based on the convolution
method that was modified for the generation of correlated binary
sequences. The relation between the {\it filtering function} --
the kernel of the convolution operator -- and the pair correlator
turns out to be relatively simple. The advantage of this method is
that it requires less computation efforts to generate long
sequences.

\section{Convolution Method}
The convolution method of generation of a \emph{continuous colored
noise} $\beta(n)$ starting from a white-noise $\alpha(n)$ is based
on a linear transformation with the use of the modulation function
$G(n)$, see Refs. [ \onlinecite{IzKr99,IzMak05,Kuhl}]. The most
general form of this transformation is as follows,
\begin{equation}\label{CM-beta-def}
\beta(n)=\overline{\beta}+\sqrt{\frac{C_\beta(0)}{C_\alpha(0)}}
\sum_{n'=-\infty}^{\infty}G(n-n')[\alpha(n')-\overline{\alpha}].
\end{equation}
Here $\overline{\alpha}$ and $\overline{\beta}$ are the mean
values, and $C_\alpha(0)= \overline{\alpha^2(n)} -
(\overline{\alpha})^2$ and $C_\beta(0)= \overline{\beta^2(n)} -
(\overline{\beta})^2$ are the variances of the white and colored
noise respectively. For a homogeneous sequence $\alpha(n)$ the
generated sequence $\beta(n)$ is also homogeneous. In what
follows, we introduce the normalized pair correlator
$K_\beta(r)=C_\beta(r)/C_\beta(0)$, which is an even function of
$r$. Substituting the linear transformation (\ref{CM-beta-def})
into the correlation function $C_\beta(r)$,
\begin{equation}
\label{cor_function} C_\beta(r)= \overline{\beta(n+r) \beta(n)}-
(\overline\beta)^2=C_\beta(0) K_\beta(r)
\end{equation}
and taking into account that the sequence $\alpha(n)$ is
$\delta$-correlated, the following relation between the pair
correlator and the modulation function is readily obtained,
\begin{equation}\label{CM-KG-Rel}
K_\beta(r)=\sum_{n=-\infty}^{\infty}G(n)G(n+r).
\end{equation}
From the condition $K_\beta(0)=1$ one gets,
\begin{equation}\label{CM-GNorm}
\sum_{n=-\infty}^{\infty}G^2(n)=1.
\end{equation}
Calculating the product $\overline{\beta(n+r) \beta(n)}$, we took
into account that the white noise $\alpha(n)$ is ergodic, i.e. the
average over $n$ can be replaced by the average over ensemble of
white-noise sequences.

Since $K_\beta(n)$ and $G(n)$ are even functions, it is convenient
to apply the cosine Fourier transform to both sides of
Eq.~(\ref{CM-KG-Rel}). This results in the following relation,
\begin{equation}\label{CM-KG-FT}
\mathcal{K}_\beta(k)=\mathcal{G}^2(k)
\end{equation}
where
\begin{eqnarray}\label{FTr-K}
\label{FTr-Kr}
\mathcal{K}_\beta(k)=1+2\sum_{r=1}^{\infty}K_\beta(r)\cos(kr),
\,\,\,K_\beta(r)=\frac{1}{\pi}\int_{0}^{\pi}
\mathcal{K}_\beta(k)\cos(kr)dk .
\end{eqnarray}
Similar relations can be written for $\mathcal{G}(k)$ and $G(n)$.

The expression (\ref{CM-KG-FT}) determines the modulation function
in terms of the Fourier transform of the pair correlator,
\begin{equation}\label{CM-G}
G(n)=\frac{1}{\pi}\int_{0}^{\pi}
\mathcal{K}_\beta^{1/2}(k)\cos(kn)dk.
\end{equation}
Evidently, the solution (\ref{CM-G}) satisfies the normalization
condition (\ref{CM-GNorm}).

For different white-noise sequences $\alpha(n)$ the convolution
method, Eqs.~(\ref{CM-beta-def}), (\ref{FTr-K}) and (\ref{CM-G}),
defines an ensemble of colored-noise sequences $\beta(n)$,
possessing the same pair correlator $K_{\beta}(n)$. The number of
terms contributing to the series (\ref{CM-beta-def}) depends on
the sharpness of the correlator $K_{\beta}(n)$. For short-range
correlations, when $K_{\beta}(n)$ decays very fast, the modulation
function $G(n)$ is also sharp, therefore, the principal
contribution is mainly given by a single term with $n^{\prime}=n$.
The sequence $\beta(n)$ in this case is practically
delta-correlated for both continuous and binary sequences
$\alpha(n)$. In the opposite case of long-range correlations, when
the correlation length $R_c$ is large ($R_c\gg 1$), many terms
contribute to Eq.~(\ref{CM-beta-def}). In this case even for a
binary sequence $\alpha(n)$, employing the method of
characteristic function, one can obtain that the probability
density $\rho_{B}(\beta)$ for stochastic variable $\beta(n)$ has
the Gaussian form,
\begin{equation}\label{CM-rhoB-fin}
\rho_{B}(\beta)=\frac{1}{\sqrt{2\pi C_{\beta}(0)}}\,
\exp\left[-\frac{(\beta-\overline{\beta})^{2}}{2C_{\beta}(0)}\right],
\end{equation}
provided the condition
\begin{equation}
\label{condition} (\beta-\overline{\beta})^{2}/2C_{\beta}(0) \ll
R_c
\end{equation}
is fulfilled. The deviations from the Gaussian shape may appear
only at the far tails, where
$(\beta-\overline{\beta})^{2}/2C_{\beta}(0)\gg R_c$. Note that for
a continuous Gaussian distribution of $\alpha(n)$
Eq.(\ref{CM-rhoB-fin}) is exact for any value of $R_c$.

From the above consideration it is clear that the correlated
sequence $\beta(n)$ may be generated using very different
uncorrelated sequences $\alpha(n)$, including the binary white
noise. However, if we start with the binary white noise
$\alpha(n)$, the generated sequence $\beta(n)$ is obviously a
non-binary one since at any site $n$ the value of $\beta(n)$ in
Eq.~(\ref{CM-beta-def}) results from a linear superposition of
binary entries. This means that the direct application of the
convolution method does not generate a binary correlated sequence.

\section{Filtering Probability}
Let us now consider the problem of generating a binary sequence
$\varepsilon(n)$ with prescribed correlations, assuming that the
sequence $\alpha(n)$ is also binary. We suppose that both the
sequence $\varepsilon(n)$ and $\alpha(n)$ contain 0's and 1's. Let
the $n$th site for $\varepsilon(n)$ is associated with the number
$P_n$ ($0\leq P_n \leq 1$), which is the probability of "1"
obtaining at this site. In order to calculate the {\it filtering
probabilities} $P_n$ from white noise $\alpha(n)$, in analogy with
Eq.~(\ref{CM-beta-def}) we propose the linear transformation
\begin{equation}\label{FPF-def}
P_n=\overline{\varepsilon}+
\sqrt{\frac{C_\varepsilon(0)}{C_\alpha(0)}}
\sum_{n'=-\infty}^{\infty}F(n-n')[\alpha(n')-\overline{\alpha}]\,.
\end{equation}
Here
$C_\varepsilon(0)=\overline{\varepsilon}(1-\overline{\varepsilon})$
and $C_\alpha(0)=\overline{\alpha}(1-\overline{\alpha})$ are the
variances of $\varepsilon(n)$ and $\alpha(n)$, respectively, and
$F(n)$ is an unknown modulation function to be determined. Having
the value of $P_n$, the $n$th symbol is generated by drawing
randomly a number from the interval [0,1]. If this number is less
than $P_n$, then $\varepsilon(n)=1$, otherwise,
$\varepsilon(n)=0$. Thus, a binary sequence $\varepsilon(n)$ can
be generated, once the set of numbers
$\left\{P_n\right\}_{n=-{\infty}}^{\infty}$ is known due to Eq.
(\ref{FPF-def}). The values of $P_n$ are correlated in the
following way,
\begin{equation}\label{F-CorFP}
\overline{[P_{n+r}-\overline{\varepsilon}]
[P_n-\overline{\varepsilon}]}
=C_\varepsilon(0)\sum_{n=-\infty}^{\infty}F(n)F(n+r),
\qquad\mbox{for}\quad r\neq0.
\end{equation}

According to the method of generation of the sequence
$\varepsilon(n)$  from the filtering probabilities, the
probability of the symbol $\varepsilon(n+r)$ obtaining at the
$(n+r)$th site does not depend on the emergence of the symbol
$\varepsilon(n)$ at $n$th site (for $r\neq0$).  Therefore, the
product $P_{n+r}P_n$ gives the joint probability of 1 obtaining at
the $n$th and $(n+r)$th sites. If 0 appears at either of these
sites, the corresponding pair does not contribute to the product
$\overline{\varepsilon(n+r)\varepsilon(n)}$. Hence, the
correlation function for sequence $\varepsilon(n)$ can be
expressed through the correlation function of the filtering
probabilities
\begin{equation}
\label{corr}
\overline{[\varepsilon(n+r)-\bar{\varepsilon}][\varepsilon(n) -
\bar{\varepsilon}]} = \lim_{N\rightarrow \infty}
\frac{1}{2N+1}\sum_{n=-N}^{N}P_{n+r}P_{n} -\bar{\varepsilon}^2=
\overline{[P_{n+r}-\overline{\varepsilon}]
[P_{n}-\overline{\varepsilon}]}.
\end{equation}
As one can see, the correlations in the sequence $\varepsilon(n)$
occur because of the correlations between the filtering
probabilities. The latter are enforced by the modulation function,
see Eq. (\ref{FPF-def}). Thus, using Eqs. (\ref{F-CorFP}) and
(\ref{corr}), the relation between the correlator of the binary
sequence and the modulation function can be written as
\begin{eqnarray}
\label{F-CorF-Rel}
&&K_\varepsilon(r)=\sum_{n=-\infty}^{\infty}F(n)F(n+r)
\qquad\mbox{for}\quad r\neq0\,;\label{F-CorF-eq}\\[6pt]
&&K_\varepsilon(0)=1. \label{F-CorNorm}
\end{eqnarray}

The normalization condition (\ref{F-CorNorm}) is the property of
the correlator $K_{\varepsilon}(r)$. It should be stressed that
unlike Eq.~(\ref{CM-KG-Rel}) that is valid for all values of $r$
including $r=0$, in the case of binary sequences the derived
Eq.~(\ref{F-CorF-Rel}) is not valid for $r=0$. Therefore, the sum
$\sum_nF^2(n)$ remains undefined and has to be considered as a
free constant,
\begin{equation}
A=\sum_{n=-\infty}^{\infty}F^2(n)
=\frac{1}{2\pi}\int_{-\pi}^{\pi}\mathcal{F}^2(k) dk.
\label{F-CorF-Adef}
\end{equation}
This constant appears now in the Fourier transform of
Eq.~(\ref{F-CorF-Rel}) as follows,
\begin{equation}
\label{F-CorF-FTRel}
\mathcal{K}_\varepsilon(k)=1-A+\mathcal{F}^2(k)
\end{equation}
Using Eq.~(\ref{F-CorF-FTRel}) the relation between the modulation
function and the pair correlator of the binary sequence can be
written in the form,
\begin{equation}\label{F-Fr-fin}
F(n)=\frac{1}{\pi}\int_{0}^{\pi}
\left[\mathcal{K}_\varepsilon(k)-1+A\right]^{1/2}\cos(kn)dk.
\end{equation}
Thus, Eqs.~(\ref{FPF-def}) and (\ref{F-Fr-fin}) define the
algorithm of the generation of a binary sequence with the
prescribed pair correlator.

The unknown constant $A$ has to satisfy to the condition that the
values of the filtering probabilities (\ref{FPF-def}) do not
exceed 1,
\begin{equation}\label{FF-restr}
\sum_{n=-\infty}^{\infty}|F(n)|\leq
\frac{\texttt{min}(\overline{\varepsilon},1-\overline{\varepsilon})}
{\sqrt{\overline{\varepsilon}(1-\overline{\varepsilon})}}\,
\frac{\sqrt{\overline{\alpha}(1-\overline{\alpha})}}
{\texttt{max}(\overline{\alpha},1-\overline{\alpha})} \leq 1.
\end{equation}
Taking into account that the argument of the square-root in Eq.
(\ref{F-Fr-fin}) must be  positive, this inequality can be
rewritten  (for $\bar\alpha=\bar\varepsilon = 1/2$) in terms of
the constant $A$,
\begin{equation}\label{FF-weakrestr-WN}
1-\mathcal{K}_\varepsilon(k)\leq
A=\sum_{n=-\infty}^{\infty}F^2(n)\leq
\sum_{n=-\infty}^{\infty}|F(n)|\leq1\,.
\end{equation}
The constraints imposed by Eqs.~(\ref{FF-restr}) and
(\ref{FF-weakrestr-WN}) limit the class of generated binary
sequences that can have a given pair correlator. In particular, a
binary sequence with the slow-decaying correlator
$K_{\varepsilon}(n) = \sin(an)/an$ cannot be generated with the
use of the proposed algorithm since the sum $\sum_{n}|F(n)|$
diverges in this case. As is known, the power-decaying correlators
provide an emergence of a kind of mobility edges in systems with
random one-dimensional potentials.~\cite{IzKr99,Lyra} In this
case, a sharp transition from localized to delocalized eigenstates
occurs when crossing some value in the energy spectrum, specified
by the mobility edge. Although such mobility edges were
experimentally observed for a continuous distribution of
fluctuations in artificially fabricated site-potentials with
long-range correlations,~\cite{Kuhl} it is not clear yet, if they
do exist for a correlated binary sequence. That is why the
existence of a mobility edge in a sequence of nucleotides in a DNA
molecule is still
questionable.~\cite{CarpBerIvStanley-Nat02,Shultz}

As an example, let us consider the exponential binary correlator
with the corresponding Fourier representation,
\begin{equation}
\label{example} K_\varepsilon(r)=\exp(-\gamma
|r|),\,\,\,\,\mathcal{K}_\varepsilon(k)=\frac{
\sinh\gamma}{\cosh\gamma-\cos k}.
\end{equation}
Since $\mathcal{K}_\varepsilon(k)$ reaches its minimum at $k=\pi$,
the condition $1- \mathcal{K}_\varepsilon(k) \leq A$ is satisfied
for $A=1-\mathcal{K}_\varepsilon(\pi)= 1-\tanh(\gamma/2)$.
Therefore, we have,
\begin{equation}
\mathcal{F}^2(k)= \frac{1+\cos k}{\cosh \gamma -\cos
k}\,\tanh\frac{\gamma}{2}\,,
\end{equation}
and the function $F(n)$ reads
\begin{equation}
F(n)= \frac{1}{\pi} \sqrt{ \tanh{\frac{ \gamma}{2}}} \int_0^\pi
\cos (kn) \sqrt \frac{1+\cos k}{\cosh \gamma -\cos k} \,\,d k\,.
\end{equation}
For $n \gg 1$ the modulation function decays as follows,
\begin{equation}
\label{decay} F(n) \approx \frac{(-1)^{n+1}}{2n^2 \cosh(\gamma/2)}
\sqrt{\tanh \frac{\gamma}{2}} \propto 1/n^2\,,
\end{equation}
therefore, the sum $\sum_{n}|F(n)|$ converges. It, however,
exceeds 1 for $\gamma < \gamma_{cr} \approx 1.60$, thus violating
the last inequality in Eq. (\ref{FF-weakrestr-WN}). The numerical
simulation shows that for $\gamma > \gamma_{cr}$ the method works
quite well, giving a possibility to construct binary chains of
length $10^7$ with the prescribed correlator $K(r)$ within a few
percents of accuracy for $r \le 5$.

\emph{In conclusion}, we suggest a method of filtering
probabilities to construct a binary correlated sequence from a
white noise. The proposed algorithm consists of the following
steps. First, starting from the prescribed mean value
$\overline{\varepsilon}$, variance $C_\varepsilon(0)$ and power
spectrum  $\mathcal{K}_\varepsilon(k)$, one calculates the
filtering function $F(n)$ by making use of Eq.~(\ref{F-Fr-fin}).
The next step is optimization of the value of the constant $A$
according to Eq.~(\ref{FF-weakrestr-WN}). Then, the filtering
probabilities $P_n$ are calculated from Eq.~(\ref{FPF-def}).
Finally, for any site $n$ by comparing the value of $P_n$ with a
number drawn randomly from the interval [0,1], one gets the number
0 or 1 that create the binary sequence $\beta(n)$.

F.M.I. acknowledges partial support from the CONACYT (Mexico),
grant No.~43730.




\begin{thebibliography}{20}

\bibitem{Hangi94} P. H\"{a}nggi and P. Jung, Adv. Chem. Phys.
{\bf 89}, 239 (1994).

\bibitem{Rice4454}
S.~O.~Rice, Bell Syst. Tech. J. {\bf 23}, 282 (1944); S.~O.~Rice, in
{\it Selected Papers on Noise and Stochastic Processes}, ed. by N.
Wax (Dover, New York, 1954) p.~180.

\bibitem{Saupe88Feder88}
D.~Saupe, in \textit{The Science of Fractal Images}, ed. by
H.-O.~Peitgen and D.~Saupe (Springer, New York, 1988); J.~Feder,
\textit{Fractals} (Plenum Press, New York, 1988).

\bibitem{WOD95}
C.~S.~West, K.~A.~O'Donnell, J. Opt. Soc. Am. A {\bf 12}, 390
(1995).

\bibitem{Czir95}
A.~Czirok, R.~N.~Mantegna, S.~Havlin, H.~E.~Stanley, \pre
{\bf{52}}, 446 (1995).


\bibitem{IzKr99}
F.~M.~Izrailev, A.~A.~Krokhin, \prl {\bf 82}, 4062 (1999).

\bibitem{IzMak05}
F.~M.~Izrailev, N.~M.~Makarov, J. Phys. A: Math. Gen. {\bf 38},
10613 (2005).

\bibitem{Grigo06} R.~Cakir, P.~Grigolini, and A.~A.~Krokhin, \pre {\bf 74}, 021108 (2006).

\bibitem{Klyat} See, for example, V.~I.~Klyatskin {\it Dynamics of Stochastic
Systems } (Elsivier, 2005); P.~A.~Mello and N.~Kumar {\it Quantum Transport in
Mesoscopic Systems : Complexity and Statistical Fluctuations}
(Oxford University Press, 2004).

\bibitem{MakYur89FMYu90}
N.~M.~Makarov, I.~V.~Yurkevich, {\it Zh. Eksp. Teor. Fiz.} {\bf
96} 1106 (1989) [{\it Sov. Phys. JETP} {\bf 69} 628 (1989)];
V.~D.~Freylikher, N.~M.~Makarov, I.~V.~Yurkevich, \prb {\bf 41},
8033 (1990).

\bibitem{Polge} R.~J.~Polge and H.~Stern, SOUTHEASTCON '81,
Proceedings of the Region 3 Conference and Exhibit, Huntsville,
AL, 1981, IEEE Inc., p. 164, 1981; R.~J.~Polge, AGARD
Multifunction Radar for Airborne Applications {\bf 9}, 11 (1986).

\bibitem{Noncorr} Sh. Hod and U. Keshet, \pre {\bf 70}, 015104(R)
(2004); I. Avgin, \prb {\bf 73}, 052201 (2006).

\bibitem{CarpBerIvStanley-Nat02}
P.~Carpena, P.~Bernaola-Gal\'van, P.~Ch.~Ivanov, H.~E.~Stanley,
Nature {\bf 418}, 955 (2002); Nature {\bf 421}, 764 (2003).

\bibitem{MUYaG05}
S.~S.~Melnyk, O.~V.~Usatenko, V.~A.~Yampol'skii, and V.~A.~Golick,
\pre {\bf 72}, 026140 (2005); S.~S.~Melnyk, O.~V.~Usatenko,
V.~A.~Yampol'skii, Physica A {\bf 361}, 405 (2005);
F.~M.~Izrailev, A.~A.~Krokhin, N.~M.~Makarov, S.~S.~Melnyk,
O.~V.~Usatenko, and V.~A.~Yampol'skii, Physica A {\bf 372}, 279
(2006).

\bibitem{Kuhl} U.~Kuhl, F.~Izrailev, A.~A.~Krokhin, and H.-J.
St\"ockmann, \apl {\bf 77}, 633 (2000).

\bibitem{Lyra} F.A.B.F.~de Moura and M.~L.~Lyra, \prl {\bf 81}, 3735
(1998).

\bibitem{Shultz} R.~A.~Caetano and P.~A.~Schulz, \prl {\bf 95}, 126601
(2005); A.~Sedrakyan and F.~Dom{\'\i}nguez-Adame, \prl {\bf 96},
059703 (2006); E.~Diaz, A.~Sedrakyan, D.~Sedrakyan, and
F.~Dom{\'\i}nguez-Adame, \prb {\bf 75}, 014201 (2007).

\end{thebibliography}
\end{document}